\documentclass[prc,showpacs,twocolumn]{revtex4}

\usepackage{graphicx}
\def\rec{recombination}
\def\dis{distribution}
\def\ptt{p_T}
\def\pt{$p_t$}
\def\pa{$p_a$}
\def\pb{$p_b$}
\def\e{Eq.\ }
\def\bl{\beta L}
\begin{document} 
\title 
{Hadron Correlation in Jets on the Near and Away Sides \\ of High-$p_T$ Triggers in Heavy-ion Collisions}
\author
 {Rudolph C. Hwa$^1$ and C.\ B.\ Yang$^{1,2}$}
\affiliation
{$^1$Institute of Theoretical Science and Department of
Physics, University of Oregon, Eugene, OR 97403-5203, USA\\
$^2$Institute of Particle Physics, Hua-Zhong Normal
University, Wuhan 430079, P.\ R.\ China}

\begin{abstract} 
Correlation between trigger and associated particles in jets produced on near and away sides of high-$p_T$ triggers in heavy-ion collisions is studied. Hadronization of jets on both sides is treated by thermal-shower and shower-shower recombination. Energy loss of semihard and hard partons traversing the nuclear medium is parametrized in a way that renders good fit of the single-particle inclusive \dis s at all centralities. The associated hadron \dis\ in the near-side jet can be determined showing weak dependence on system size because of trigger bias. The inverse slope increases with trigger momentum in agreement with data. The \dis\ of associated particle in the away-side jet is also studied with careful attention given to antitrigger bias that is due to the longer path length that the away-side jet recoiling against the trigger jet must propagate in the medium to reach the opposite side. Centrality dependence is taken into account after determining a realistic probability distribution of the dynamical path length of  the parton trajectory within each class of centrality. For symmetric dijets with $p_T^{\rm trig}=p_T^{\rm assoc}({\rm away})$ it is shown that the per-trigger yield is dominated by tangential jets. For unequal $p_T^{\rm trig},\  p_T^{\rm assoc}({\rm near})$ and $p_T^{\rm assoc}({\rm away})$, the yields are calculated for various centralities, showing intricate relationship among them.  The near-side yield agrees with data both in centrality dependence and in $p_T^{\rm assoc}({\rm near})$ distribution. The average parton momentum for the recoil jet is shown to be always larger than that of the trigger jet for fixed $p_T^{\rm trig}$ and centrality and for any measurable $p_T^{\rm assoc}({\rm away})$. With  the comprehensive treatment of dijet production described here it is possible to answer many questions regarding the behavior of partons in the medium under conditions that can be specified on measurable hadron momenta.

 \end{abstract}
\maketitle
\section{Introduction}

Recent studies of jet correlation in heavy-ion collisions at RHIC have generated a wealth of information about jet-medium interaction, not only on how the dense medium modifies the characteristics of high-$p_T$ jets, but also on how intermediate-$p_T$ jets affect the medium \cite{ca,ja,aa,aa2}.  Two-particle correlation has been particularly effective in revealing the nature of the medium response to the passage of a hard or semihard parton \cite{rev}.  The discovery of ridge, for example, in the structure of the same-side \dis\ of particles associated with a trigger has stimulated intense interest both experimentally and theoretically \cite{jp}-\cite{af}.  The properties of the ridge (in centrality and $p_T$ dependencies, and in baryon/meson ratio) distinguish its origin from that of the jet peak that stands above the ridge.  Similar distinction can be found between the punch-through jet and the double-hump peaks on the away side.  In this paper we study the properties of the associated jets on both sides.  We calculate not only the $\ptt$ \dis s of the particles in those jets, but also the fractional energy loss of the hard partons traversing the medium toward and away from the trigger. As a consequence we can
quantify the notion of trigger and antitrigger biases.

It has come to be generally accepted that the hadronization process at intermediate $p_T$ is \rec /coalescene \cite{hy,gkl,fmnb}.  The approach that we have adopted in Ref.\  \cite{hy2} emphasizes the role that shower partons play in interpolating the production processes from thermal-thermal (TT) \rec\ at low $p_T$ to shower-shower (SS) \rec\ at high $p_T$, which is identical to fragmentation, through the intermediate region where thermal-shower (TS) \rec\ is important.  The application of that approach to dihadron correlation has been considered previously \cite{hy3,ht}, but before the discovery of ridges.  The phenomenology of ridges (or ridgeology) has clarified the characteristics of the associated particles on the near side.  There are strong indications that the ridge particles are formed by the \rec\ of enhanced thermal partons \cite{rch}-\cite{ch3}.  Thus after subtracting out the ridge particles, what remain are the jet particles, which being close to the trigger in $\Delta\phi$, are due exclusively to TS and SS \rec.  With the experimental data on the jets being refined, it is now appropriate to reexamine the jet correlation problem in both the near and away sides.

Our formalism allows us to study the trigger bias on the near side and the antitrigger bias on the away side, i.e., higher average initiating parton momentum in order to allow for more energy loss in traversing longer path length to reach the other side.    The average transverse momentum  $\left< p_T\right>$ on either side, which is related to the inverse slope, turns out to depend sensitively on the origin of the partons that hadronize.  TS \rec\ has a softer $p_T$ \dis\ than SS \rec\ (or fragmentation).  The varying mixture of TS and SS components in different $p_T$ ranges makes the hadronization process an integral part of any procedure to associate the characteristics of the $p_T$ \dis\ with either the trigger or antitrigger effect.  We are able to calculate the inverse slopes of the associated particles on both the near and away sides. Data exist for the former, since the ridge contribution has been studied experimentally in detail and can be subtracted. Our result agrees very well with those data. The associated particles in the away-side jets are hard to analyze because of the double-hump background that is difficult to separate, so our results on those jets cannot yet be checked by data. For fixed medium suppression in central collisions, we have studied various other quantities that are  not directly measurable in experiments, but they can shed considerable light on the trigger and antitrigger  effects. 

For realistic nuclear collisions the path lengths of hard partons and the quenching effect depend on the point of hard scattering in the transverse plane and the azimuthal angle of the trajectory.  An important part of our study is to find a way to describe the variation of both the path length and the quenching effect, what may be called the dynamical path length, for different collisions within each class of centrality. A \dis\ of that measure will play a crucial role in relating theoretical calculations to experimental observation at definite intervals of centrality. Such a \dis\ has been found in our study and is shown to render an excellent reproduction of the dependence of the inclusive spectra of pions in the range $2<\ptt<11$ GeV/c at all centralities. With that \dis\ at hand, we can then calculate dihadron correlation that can realistically describe properties of dijet production.
 We find that the dominance of tangential jets emerges naturally as the momenta of trigger and associated particle on opposite sides approach each other. It is a clear demonstration of the interplay between trigger and antitrigger effects.

\section{Dihadron Correlation in the Recombination Model}

We adopt the formalism initiated in \cite{hy2} for single-particle inclusive \dis\ and in \cite{hy3} for dihadron correlation. Concentrating on only the jet component of the associated particles, we ignore TT \rec\, which gives rise to the ridge on the near side and to the double-hump on the away side.  The medium effect is parameterized by an exponential damping factor that depends on the path length.  Our focus is on the momentum \dis\ of the associated particles for trigger-particle momentum larger than 4 GeV/c.  We restrict our consideration to midrapidity and study only the transverse momentum $p_T$.  Thus we shall omit the subscript $T$, and use $p_t$ to denote trigger momentum, $p_a$ for associated particle on the near side and $p_b$ for associated particle on the away side.  Without any subscript, $p$ shall be used as a generic symbol for the transverse momentum of any hadron.

In the simplest form the invariant \dis\  of a pion is \cite{hy2}
\begin{eqnarray}
p  {dN_{\pi}\over dp} = \int {dq_1 \over q_1}{dq_2 \over q_2} F_{q\bar{q}}(q_1, q_2)R_{\pi}(q_1, q_2,p)
\label{1}
\end{eqnarray}
where the $q\bar{q}$ \dis\ is in general
\begin{eqnarray}
F_{q\bar{q}} (q_1, q_2) = {\cal TT} + {\cal TS} + {\cal SS}
\label{2}
\end{eqnarray}
and the recombination function (RF)
\begin{eqnarray}
R_{\pi}(q_1, q_2,p) = {q_1q_2 \over p^2} \delta \left({q_1\over p}+{q_2\over p} -1 \right) .
\label{3}
\end{eqnarray}  
The thermal parton \dis\ has the form
\begin{eqnarray}
{\cal T} (q_1) = q_1 {dN^{\rm th}_q \over dq_1} = Cq_1e ^{-q_1/T}
\label{4}
\end{eqnarray}  
so that the thermal pion \dis\ is exponential
\begin{eqnarray}
{dN^{TT}_{\pi} \over pdp} = {C^2 \over 6} e ^{-p/T} .
\label{5}
\end{eqnarray}
The quark momentum $q_i$ above are just before hadronization at the end of medium expansion.  The shower parton at that stage is in the vacuum after the hard parton has emerged from the medium.  Just after hard scattering the \dis\ of the hard parton momentum $k$ of parton type $i$, while still in the medium, is given by 
\begin{eqnarray}
\left.{dN^{{\rm hard}}_{i} \over kdkdy}\right|_{y = 0} = f_i(k) ,
\label{6}
\end{eqnarray}
whose specific properties are given in the next section.
After propagating through the medium, the parton loses momentum in a way that we describe by the  function $G(q, k, t)$ where $t$ denotes the distance the parton travels to reach the surface, and $q$  is the  momentum of that parton at the surface. We discuss $G(q, k, t)$ below presently. The parton \dis\ in $q$ after averaging over all $k$ and $t$ is
\begin{eqnarray}
F_i (q) = \int^L_0 {dt \over L} \int^{\infty}_{k_0} dk k f_i (k) G(q, k, t) ,
\label{7}
\end{eqnarray}
where in calculation  we set the lower limit $k_0$ at 3 GeV, below which the parton \dis\ $f_i(k)$ is not known reliably.  $L$ is the average maximum length of the system that the hard parton traverses. In the limit $L \to 0$, we should recover the parton distribution for $pp$ collision.

For the degradation factor $G(q, k, t)$ due to energy loss, there is a rich literature on the subject studied by various methods. Two articles reviewing the subject are Refs.\ \cite{gy,kw}. The quenching factor $Q(p)$ determined in Ref.\ \cite{bd} increases with $p$, a property at very high energy not found at RHIC. In the opacity expansion approach \cite{glv} the energy loss is found to depend on the path length as $\Delta E \propto L^{2-\alpha}$, where $\alpha=1$ for one-dimensional expansion. The effective quark energy loss with detailed balance between induced gluon emission and absorption taken into account has the form for a 1-d expanding medium \cite{xw}
\begin{eqnarray}
\left\langle {dE\over dL}\right\rangle_{1d}=\epsilon_0 (E/\mu -1.6)^{1.2}/(7.5+E/\mu) ,   \label{8}
\end{eqnarray}
which is essentially $\propto E$ for $4<E<12$ GeV. A reasonable summary of these properties is 
 \begin{eqnarray}
{\Delta E\over E}=\beta \Delta L  ,  \label{9}
\end{eqnarray}
whose implication  for the relationship between $q$ and $k$ in Eq.\ (\ref{7}) is that 
\begin{eqnarray}
k-q=k\beta t .  \label{10}
\end{eqnarray}
For $t$ not infinitesimal, we exponentiate Eq.\ (\ref{10}) and get
\begin{eqnarray}
q=k e^{-\beta t} .   \label{11}
\end{eqnarray}
Fluctuation from this relationship is undoubtedly possible, but we shall take the simple form
\begin{eqnarray}
G(q, k, t) = q \delta(q - k e^{-\beta t})   \label{12}
\end{eqnarray}
as an adequate approximation of the complicated processes involved in the parton-medium interaction. The justification for Eq.\ (\ref{12}) is to be found in the degree to which the inclusive cross section can be reproduced in our description of hadron production at intermediate and high $p_T$.

Using Eq.\ (\ref{12}), the integration over $t$ in Eq.\ (\ref{7}) can readily be carried out, giving
\begin{eqnarray}
F_i (q) = {1 \over \beta L}\int^{q e^{\beta L}}_{q} dk k f_i (k)
\label{13}
\end{eqnarray}
for $q> k_0$.
The lower limit of the above integration corresponds to $t=0$, i.e., when the hard scattering occurs at the surface, while the upper limit corresponds to the hard-scattering point being on the far side so that $k$ is a factor $e^{\beta L}$ larger than $q$.
   Equation (\ref{13}) exhibits the nuclear effect in changing $f_i(k)$ to $F_i (q) $ with $\beta L$ being the explicit medium factor, while $f_i (k)$ contains the hidden modification of the parton \dis s in the nucleus due to shadowing, etc. \cite{sgf} Clearly, as $\beta L \to 0$, $F_i(q)$ becomes directly related to $f_i(q)$ appropriately extrapolated to $pp$ collision.

Using $S^j_i$ to denote the matrix of shower parton \dis s (SPDs) that are calculable from the fragmentation functions \cite{hy4}, we can determine the \dis\ of shower partons in a heavy-ion collision by 
\begin{eqnarray}
{\cal S}(q_1) = \int {dq \over q} F_i (q) S^{j}_i (q_1/q).
\label{14}
\end{eqnarray}
The {\cal TS} contribution to the inclusive pion \dis\ is then, following Eqs.\ (\ref{1}), (\ref{2}), (\ref{4}) and (\ref{14}), 
\begin{eqnarray}
{dN^{TS}_{\pi} \over pdp} = {1 \over p^2} \sum_i \int {dq \over q} F_i(q)\widehat{\sf TS} (q, p) ,
\label{15}
\end{eqnarray}
with the RF absorbed in the compound notation for the TS term in the integrand:
\begin{eqnarray}
\widehat{\sf TS} (q, p) = \int {dq_1 \over q_1} S^j_i \left({q_1 \over q}\right) \int dq_2 C_{\bar j} e^{-q_2/T} {R}_{j \bar{j}}(q_1, q_2, p) ,
\label{16}
\end{eqnarray}
where for every hard-scattered parton of type $i$ the shower parton of type $j$ is paired with a thermal parton of type $\bar{j}$ for recombination in forming a pion.  For the SS component we can use the fragmentation function $D(z)$ and write
\begin{eqnarray}
{dN^{SS}_{\pi} \over pdp} = {1 \over p^2} \sum_i \int {dq \over q} F_i (q) {p \over q} D^{\pi}_i \left({p\over q} \right).
\label{17}
\end{eqnarray}
The overall pion inclusive \dis, including the TT contribution as given in Eq.\ (\ref{5}), is thus
\begin{eqnarray}
{dN_{\pi} \over pdp} &=& {C^2 \over 6} e ^{-p/T}  
+ {1 \over p^2} \sum_i \int {dq \over q} F_i (q) \nonumber\\
&&\times \left[\widehat{\sf TS} (q, p)+{p \over q} D^{\pi}_i \left({p\over q} \right)\right].
\label{18}
\end{eqnarray}
  
For the dihadron correlation on the same side we consider the trigger-momentum \pt\ to be greater than 4 GeV/c and calculate the associated particle \dis\ in the approximation that the TT contribution to the trigger and jet can be neglected.  We then have for \pa\ on the near side associated with \pt\
\begin{eqnarray}
{dN_{\pi\pi} \over p_tp_a dp_t dp_a}&=& {1 \over (p_tp_a)^2} \sum_i \int {dq \over q} F_i (q) \nonumber\\
&&\times \left\{\left[\widehat{\sf TS}  (q,p_t) + {p_t \over q} D^{\pi}_i  \left({p_t \over q}\right)\right]\widehat{\sf TS}  (q - p_t,p_a) \right.  \nonumber\\
&&+ \widehat{\sf TS}  (q - p_a,p_t){p_a \over q}D^{\pi}_i  \left({p_a \over q}\right) 
\nonumber\\ &&
\left. + {p_t p_a \over q^2_j }D^{\pi}_2  \left( {p_t \over q},  {p_a \over q}\right)\right\}
\label{19}
\end{eqnarray}
where the dihadron fragmentation function $D_2(z_1, z_2)$ is assumed to have the symmetrized form
\begin{eqnarray}
D_2(z_1, z_2) = &&{1 \over 2} \left[ D(z_1) D\left( {z_2 \over 1 - z_1}\right)\right. \nonumber\\
&&\left.+ D\left( {z_1 \over 1 - z_2}\right) D(z_2)\right] .
\label{20}
\end{eqnarray}
The near-side yield per trigger for trigger momentum in a narrow range $\Delta p_t$ around \pt\ is
\begin{eqnarray}
Y_{\pi\pi}^{\rm near}(p_t, p_a)&=&{1\over N_{\rm trig}}{dN_{\pi\pi}\over p_adp_a}(p_t,p_a) \nonumber\\
&=&\int_{\Delta p_t}dp_t{dN_{\pi\pi}\over p_adp_tdp_a}\left/\int_{\Delta p_t}dp_t{dN_{\pi}\over dp_t}\right., \qquad \label{21}
\end{eqnarray}
where $dN_{\pi}/dp_t$ is the trigger pion \dis\ that excludes the TT component of the inclusive \dis\ given in Eq.\ (\ref{18}).

For an associated particle on the away side relative to the trigger we must consider the recoil hard parton that propagates a distance $L - t$ in the opposite direction, so Eq.\  (\ref{7}) should be revised to contain another parton with momentum $q'$ exiting on the away side, having the form
\begin{eqnarray}
F'_i (q,q') &=& \int ^L_0 {dt \over L} \int^{\infty}_{k_0} dk k f_i (k) G (q, k, t) G (q', k,L- t)\nonumber\\
&=&{1 \over \beta L} \int^{qe^{\beta L}}_{q} dk k f_i (k) q q' \delta(q q' - k^2e^{-\beta L}) ,
\label{22}
\end{eqnarray}
where the recoil parton has momentum $k$ under the assumption of negligible initial $k_T$ of the beam partons. However, for reasons that will become clear later we label it by $k'$ to be distinguished from $k$ of the trigger parton, when clarity is needed. To help with the visualization of the various momentum variables in the problem, we give a sketch in Fig.\ 1 of the parton (red) and hadron (blue) momentum vectors with the trigger being on the right side.  With due caution in the interpretation of momentum vectors drawn in coordinate space, momenta on the near side ($q, p_t$ and \pa) are depicted to originate from the surface on the right, while  the momenta on the  away side ($q'$ and \pb) are pointed from the surface on the left.

\begin{figure}[tbph]
\includegraphics[width=0.45\textwidth,
clip=]{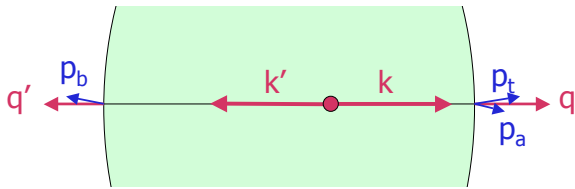}
\caption{(Color online) A sketch of momentum vectors of partons (in red) and hadrons (in blue) with near side being on the right and away side on the left.}
\end{figure}

The recoil parton of type $i'$ need not be linked to the parton type $i$ of the trigger jet, since beam partons can be of any type. However, to limit the problem to a manageable size we make the simplifying assumption that $i'=\bar i$ if they are quarks and $i'=i$ if they are gluons (thus tantamount to assuming dominance by $g+g$ scattering), and we shall only calculate identical pions on the two sides.
The dipion \dis\ for particle with \pb\ on the away side is thus
\begin{eqnarray}
{dN_{\pi\pi} \over p_t p_b dp_t dp_b} &=& {1 \over (p_t p_b)^2} \sum_i \int {dq \over q} {dq' \over q'}F'_i (q,q') \nonumber\\
&&\times \left[\widehat{\sf TS}  (q,p_t) + {p_t \over q} D^{\pi}_i\left({p_t \over q}\right)\right] \nonumber\\
&&\times  \left[  \widehat{\sf TS}  (q', p_b) + {p_b \over q'}D^{\pi}_{i'}\left({p_b \over q'}\right)   \right] .
\label{23}
\end{eqnarray}
Carrying out the integration over $k$ yields
\begin{eqnarray}
{dN_{\pi\pi} \over p_t p_b dp_t dp_b} &=& {e^{\beta L} \over 2\beta L p_t^2 p_b^2} \sum_i \int_{p_t} dq \int_{q'_0}^{qe^{\beta L}}dq' f_i \left(\sqrt{qq'e^{\beta L}}\right) \nonumber\\
&&\times \left[\widehat{\sf TS}  (q,p_t) + {p_t \over q} D^{\pi}_i\left({p_t \over q}\right)\right] \nonumber\\
&&\times  \left[  \widehat{\sf TS}  (q', p_b) + {p_b \over q'}D^{\pi}_{i'}\left({p_b \over q'}\right)   \right] ,
\label{24}
\end{eqnarray}
where $q'_0={\rm Max}(qe^{-\beta L}, p_b)$.

We note that the limits of integration of $q'$ in Eq.\ (\ref{24}) reveals the medium effect  in the following sense. If the hard scattering occurs at the near-side surface, then the recoil parton (having $k'=k=q$) must travel a distance $L$ to emerge on the away side with momentum $q'=q'_0=qe^{-\beta L}$. On the other hand, if the hard scattering occurs at the away-side surface, then $q$ must be $ke^{-\beta L}$, so $q'=k=qe^{\beta L}$. Thus the integration over $q'$ reflects the integration over all points $t$ in the medium where the hard parton is created.

The away-side yield per trigger is, analogous to Eq.\ (\ref{21}),
\begin{eqnarray}
Y_{\pi\pi}^{\rm away}(p_t, p_b)&=&{1\over N_{\rm trig}}{dN_{\pi\pi}\over p_bdp_b}(p_t,p_b)\nonumber\\
&=&\int_{\Delta p_t}dp_t{dN_{\pi\pi}\over p_bdp_tdp_b}\left/\int_{\Delta p_t}dp_t{dN_{\pi}\over dp_t}\right. .  \label{25}
\end{eqnarray}
In our calculation we can, of course, take the theoretical limit $\Delta p_t\to 0$.

\section{Model Inputs}

We list here all the inputs to the model that we use to perform our calculation. They are all taken from previous work without any parameters to adjust, except for $\beta L$
 that is introduced here in lieu of an average suppression factor used earlier. 
 
 The hard-scattered parton \dis\ $f_i(k)$ is taken from Ref.\ \cite{sgf}, which uses the parametrization
 \begin{eqnarray}
f_i(k)=K{A\over (1+k/B)^a}   \label{26}
\end{eqnarray}
with $K=2.5$ and the parameters $A, B$, and $a$ tabulated for each parton type $i$ and for Au+Au collisions at RHIC with shadowing taken into account. The sum $\sum_i$ will be performed over $i=g, u, d, s, \bar u, \bar d, \bar s$. For the thermal partons the values of $C$ and $T$ in Eq.\ (\ref{5}) for 0-10\% centrality are \cite{hy2}
\begin{eqnarray}
C=23.2\ {\rm GeV}^{-1} ,   \quad T=0.317\
 {\rm GeV}.   \label{27}
\end{eqnarray}
Their centrality dependence are given in Ref.\ \cite{ht}. The shower parton \dis s are described in Refs.\ \cite{hy2, hy4}. For $z$ in $S_i^j(z)$ very small, the \dis s are not reliable, so we cut off the low-$p_T$ contribution to the TS component by a factor $1-\exp(-0.5p_T)$, which has no effect on our result for intermediate and high $p_T$. For the fragmentation function $D(z)$ we use the parametrization in Ref.\ \cite{bkk}, from which the shower parton \dis s were derived \cite{hy4}. Since we now consider higher $p_T$ than before, the $Q^2$ dependence of $D(z, Q^2)$ will be included by setting $Q^2=p_T^2$.

With these inputs specified there are no more free parameters to adjust, except the suppression factor quantified by $\beta L$, which will be determined below by fitting the overall single-pion inclusive \dis. The properties of the dihadron correlations on both the near  and  away sides can then be calculated without unknown parameters.

\section{Near-side Correlation and Trigger Bias}

We first calculate the pion inclusive \dis\ using Eq.\ (\ref{18}) and compare the result to the data in Fig.\ 2. What we have calculated is $dN_{\pi}/p_Tdp_T$ at midrapidity averaged over all $\phi$, while the data are for $dN_{\pi^0}/2\pi p_Tdp_T$ integrated over all $\phi$ \cite{phe}, both for 0-10\% centrality in Au+Au collision at $\sqrt s=200$ GeV. The value 
\begin{eqnarray}
\beta L=2.9  \label{28}
\end{eqnarray}
has been used to fit the data for $2<p_T<13$ GeV/c. Since the suppression factor involving $\beta L$ enters Eq.\ (\ref{18}) only through $F_i(q)$ given in Eq.\ (\ref{13}), the excellent fit in Fig.\ 2 over such a wide range of $\ptt$ requires a high degree of coordination among the three components of \rec, and therefore is not a trivial result from varying one quantity, $\beta L$. The agreement with data is a confirmation of the soundness of the model for the range of $\ptt$ considered.

\begin{figure}[tbph]
\includegraphics[width=0.45\textwidth]{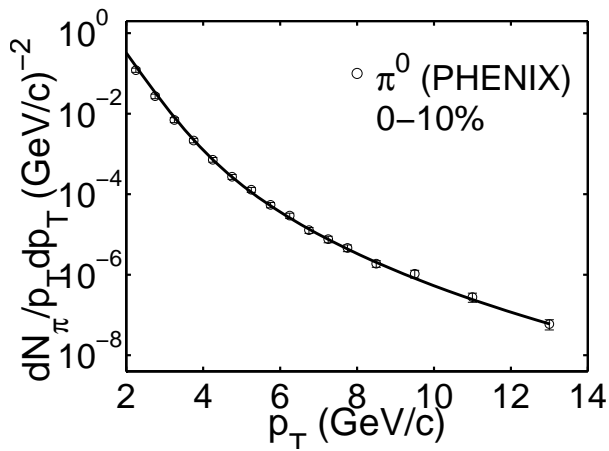}
\caption{Inclusive \dis\ of $\pi^0$ as calculated from Eq.\ (\ref{18}). The data are from Ref.\ \cite{phe}.}
\end{figure}
\begin{figure}[tbph]
\includegraphics[width=0.5\textwidth]{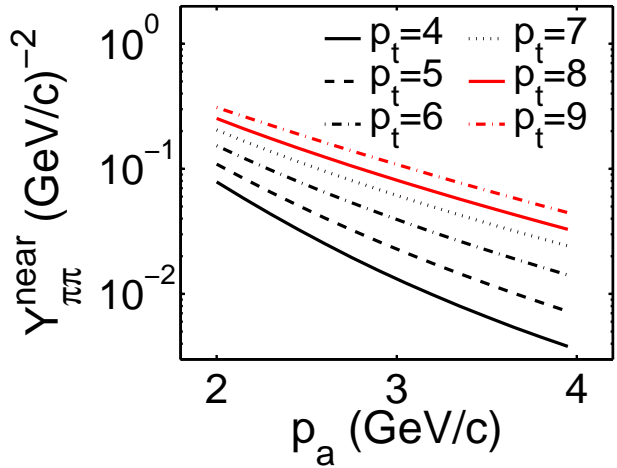}
\caption{(Color online)
Distribution of associated pion (\pa) in near-side jet for six values of pion trigger momentum (\pt) in GeV/c.}
\end{figure}

Using $\beta L$ given in Eq.\ (\ref{28}), we can now calculate the dihadron correlation on the near side for trigger momentum at $p_t>4$ GeV/c and an associated particle in the jet with momentum in the range $2<p_a<4$ GeV/c. We use \e (\ref{19}) to study the $\pi\pi$ correlation without TT \rec, which we have assigned to the ridge. The result for the yield per trigger is shown in Fig.\ 3 for 6 values of \pt. It is evident that the \pa\ spectrum becomes slightly harder, as \pt\ increases. The effective inverse slope $T_a$ determined in the range $2<p_a<4$ GeV/c is shown in Fig.\ 4. The data in that figure are from Refs.\ \cite{jp, jb} for all charged hadrons. Although our result is for pions only, we expect that the contributions from the other charged hadrons are not as important in the jet as they are in the ridge. Thus the general agreement of our result with the data may be regarded as  supportive of our description of the physics that generates the dihadron correlation.

\begin{figure}[tbph]
\includegraphics[width=0.45\textwidth]{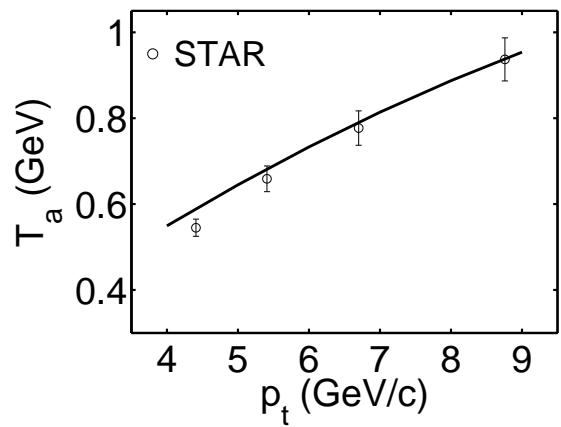}
\caption{Inverse slope of associated pion in near-side jet. Data are from Refs.\ \cite{jp, jb} determined from $2<p_T^{\rm assoc}<4$ GeV/c.}
\end{figure}

The results obtained so far average over all possible points of hard scattering, as indicated by the integration over $t$ in \e (\ref{7}). However, we know that with $\beta L=2.9$ the contributions from the points on the far side of the medium are more suppressed compared to those from the nearer points. That is the trigger bias in heavy-ion collision. We can quantify that effect by calculating the average of $\exp(-\beta t)$. Such an averaging process is feasible by using Eqs.\ (\ref{7}) and (\ref{18}), where we insert $\exp(-\beta t)$ in \e(\ref{7}) before integration over $t$. It is not necessary to know $\beta$ separately from $\beta L$ because of the structure of $G(q, k, t)$ that demands $e^{-\beta t}=q/k$. With the near-side suppression factor defined as
\begin{eqnarray}
\Gamma_{\rm near}(p_T)=\langle e^{-\beta t}\rangle,  \label{29} 
\end{eqnarray}
which is also  $\langle q/k\rangle$, we show our calculated result in Fig.\ 5. Evidently, it saturates at 0.85. Thus on average only 15\% of the parton energy is lost to the medium when $\ptt$ is high, but more at lower $\ptt$. This result is roughly independent of the medium size $L$, and is a feature of the trigger bias. That is, if the point of origin is allowed to vary, the detected hadrons are dominantly due to partons created near the surface and losing only a small fraction of the energy.

\begin{figure}[tbph]
\includegraphics[width=0.4\textwidth]{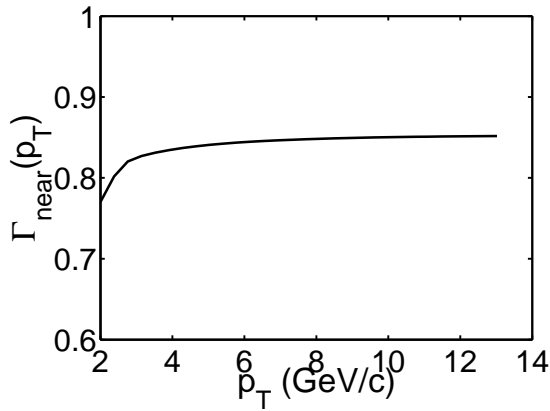}
\caption{Near-side suppression factor for which the value of 1 means no suppression. The averaging is done over single-paticle \dis\ of pion momentum $\ptt$.}
\end{figure}
\begin{figure}[tbph]
\includegraphics[width=0.45\textwidth]{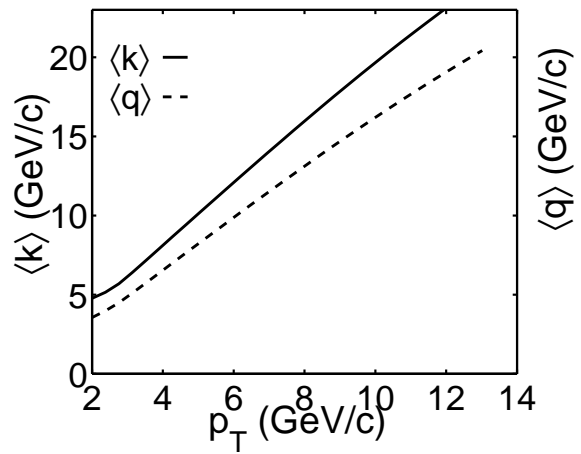}
\caption{Average values of parton momenta $k$ (at point of hard scattering) and $q$ (at the near-side surface) for pions detected at $\ptt$.}
\end{figure}

In Fig.\ 6 we show both $\langle k\rangle$ (in solid line) and $\langle q \rangle$ (in dashed line) as functions of $\ptt$. Their ratio $\langle q\rangle/\langle k\rangle$ is not exactly $\langle q/k\rangle$. The two lines provide insight into how hard the hard scattering has to be in order to give rise to a pion at $\ptt$. Note that $\langle k\rangle$ is approximately $2\, \ptt$ throughout the range, while $\langle q\rangle \approx 1.6\,  \ptt$ for
$ \ptt>3$ GeV/c where both TS and SS components of \rec\ are important. 
Since $\langle \exp(-\beta t)\rangle \ne \exp\langle -\beta t\rangle$, we have calculated $\langle \beta t\rangle$ shown in Fig.\ 7. It is an estimate of $\Delta E/E$ according to Eq.\ (\ref{9}), although TS \rec\ renders the connection with parton energy loss imprecise. For $\ptt>4$ GeV/c the average $\langle \beta t\rangle$ is between 0.18 and 0.2. That is to be compared to $\beta L=2.9$, implying $\langle t\rangle/L \approx 0.065$. Thus the result suggests that the thickness of the layer near the surface where hard partons are created is roughly 13\% of $L$. That is a quantitative statement about the trigger bias to the extent that we can calculate without taking into account such details as nuclear geometry, tangential jets, etc.

\begin{figure}[tbph]
\includegraphics[width=0.45\textwidth]{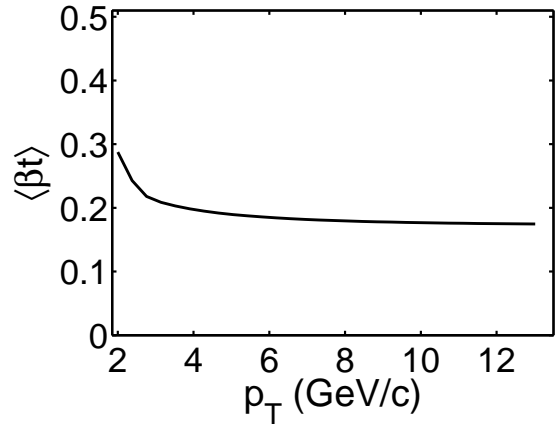}
\caption{Average value of $\beta t$ as a function of pion momentum $\ptt$.}
\end{figure}

\section{Away-side Correlation and Antitrigger Bias}

As we consider the correlation between jets on opposite sides, we first fix $\beta L=2.9$, which corresponds to a slab of nuclear medium with fixed thickness. That is not a realistic nuclear medium, whose thickness depends on the transverse distance from the center. The value of $\beta L$ determined in Eq.\ (\ref{28}) corresponds to the effective thickness in fitting the single-particle \dis\ shown in Fig.\ 2. Due to trigger bias the correlation between particles on the same side is mostly independent of that thickness, as we have seen in the preceding section. Now, as we go to dijet correlation on the two sides, it makes a big difference whether the medium has varying thickness. In order to illuminate the nature of the away-side correlation and antitrigger bias, we first consider  in this section the simplest scenario of fixed thickness. After becoming familiar with the issues involved, we then extend our study to the case of realistic nuclear medium  in the next section.

\begin{figure}[tbph]
\includegraphics[width=0.5\textwidth]{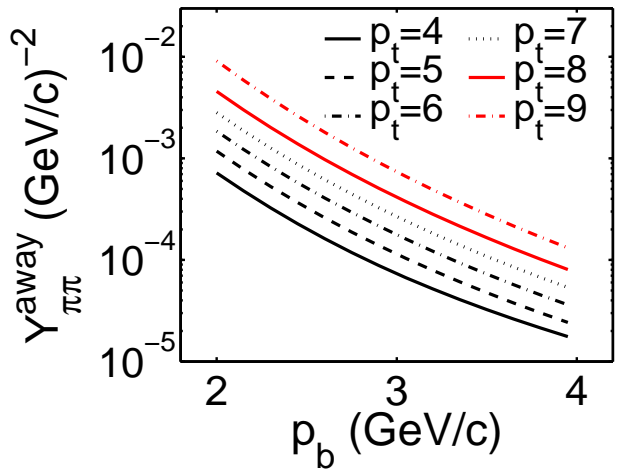}
\caption{(Color online) Distribution of associated pion (\pb) in the away-side jet for six values of pion trigger momentum (\pt) in GeV/c.}
\end{figure}
\begin{figure}[tbph]
\includegraphics[width=0.4\textwidth]{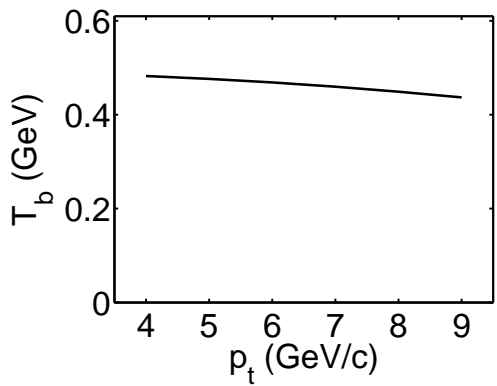}
\caption{Inverse slope of associated pion in away-side jet.}
\end{figure}

For the away-side yield per trigger we use Eqs.\ (\ref{24}) and (\ref{25}) to calculate $Y_{\pi\pi}^{\rm away}(p_t, p_b)$ as functions of \pb\ for six values of \pt. The results are shown in Fig.\ 8. As with near-side yield, the away-side per-trigger yield increases with \pt, not simply because the corresponding hard parton $k$ is forced to be higher, but also because the number of triggers is lower.
However, the spectrum does not becomes harder at larger \pt, for a reason to be discussed later. Figure 9 shows the inverse slope $T_b$ determined in the range $2<p_b<4$ GeV/c, exhibiting only a mild decrease of $\sim 10$\% over the range of \pt.
Since it is a property of the jet yield, there are no suitable data to compare with our result. PHENIX has extensive data on dihadron correlation \cite{aa2}; however, on the away side the division between head and shoulder regions is done in terms of cuts in $\Delta\phi$, with the consequence that a direct relationship between the yield in the head region and the jet yield calculated here cannot easily be established. Data on inclusive $\gamma$ have been analyzed for correlated hadrons, using a Gaussian description for punch-through jets on the away side \cite{chc}, but no $p_T$ \dis\ has been shown. 

To learn about the medium effect on the away-side jet, we study the suppression factor defined as
\begin{eqnarray}
\Gamma_{\rm away}(p_t, p_b)=\langle \exp[-\beta(L-t)]\rangle.  \label{30}
\end{eqnarray}
It should be recognized that whereas $\Gamma_{\rm near}(p_T)$ involves an average over the single-particle \dis, $\Gamma_{\rm away}(p_t,p_b)$ requires for the averaging process the opposite-side dihadron correlation that depends on \pt\ and \pb.
Our results are shown in Fig.\ 10 as functions of \pb\ for four values of \pt. As expected, the suppression is far more severe on the away side than on the near side. The higher the trigger momentum \pt, the closer is the hard-scattering point to the surface on the near side due to trigger bias, and we see that the more severe is the suppression on the away side. That is a property of antitrigger bias because of the longer path length that the recoil parton must travel in the medium. Whereas $\Gamma_{\rm near}(p_T)$ increases with $p_T$, here $\Gamma_{\rm away}(p_t,p_b)$ decreases with \pt, but increases with \pb\ for a fixed \pt. That is because higher \pb\ requires higher hard-parton $k$ or shorter $L-t$. Fixing \pt\ does not fix $k$, as the value of $\langle k\rangle$ in Fig.\ 6 is determined without any extra condition. Now, with higher \pb, $k$ must be higher, as well as $t$ must be larger, resulting in increasing $\langle \exp[-\beta(L-t)]\rangle$. In other words, the condition of higher \pb\ favors the hard-scattering points closer to the away-side jet in order to reduce suppression.

\begin{figure}[tbph]
\includegraphics[width=0.45\textwidth]{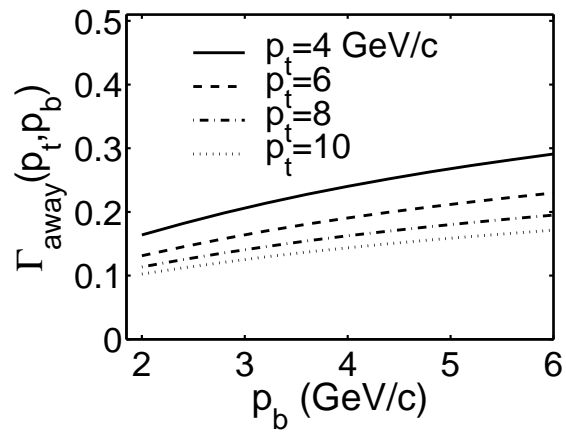}
\caption{Away-side suppression factor determined by averaging over opposite-side $\pi\pi$ correlation function with trigger momentum at \pt\ and associated-particle momentum at \pb.}
\end{figure}
\begin{figure}[tbph]
\includegraphics[width=0.45\textwidth]{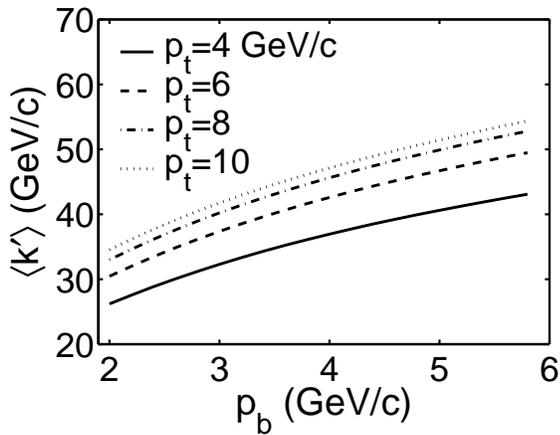}
\caption{Average value of recoil parton momentum $k'$ for various values of trigger \pt\ and associated-particle \pb\ in the away-side jet.}
\end{figure}

In a hard scattering process the two outgoing hard partons have equal and opposite momenta, if we ignore the transverse momenta $k_T$ of the initial partons. However, the averages of the two hard parton momenta may differ, depending on what observables are held fixed. That is, if $k'$ is the recoil momentum, opposite to $k$ that generates the trigger, we have $k'=k$ event-by-event, but $\langle k'\rangle(p_t,p_b)$ may well be different from $\langle k\rangle(p_T)$. We have seen how $\langle k\rangle(p_T)$ depends on $\ptt$ in Fig.\ 6. We now show in Fig.\ 11 the dependence of $\langle k'\rangle(p_t,p_b)$ on \pb\ for four values of \pt. Evidently, $\langle k'\rangle$ is much larger than $\langle k\rangle$ for  all values of \pb\ and \pt. This is the essence of antitrigger bias. The condition of having a hadron on the away side among the triggered events gives higher weight to the larger $k'$ processes in the averaging. That is why $\langle k'\rangle$ increases with both \pt\ and \pb.

As \pb\ increases, $\langle k'\rangle$ must increase in order to provide enough $\langle q'\rangle$ that can accommodate the larger \pb. Figure 12 shows $\langle q'\rangle(p_t,p_b)$, which exhibits  a mild dependence on \pt, but rises almost linearly with \pb\ above 3 GeV/c as $\sim (1.5\div 2) p_b$. That dependence on \pb\ is roughly the same as the dependence of $\langle q\rangle$ on $\ptt$ shown in Fig. 6, as it should, since the hadronization processes are similar.
The weak dependence on \pt\ implies that as \pt\ is increased, the point of hard scattering is pulled to the near side at the same time as the scattered-parton momentum $k$ is increased (see Fig.\ 6), so their opposite effects on the away side $q'$ are nearly canceled. With $\langle q'\rangle$ not changing much  in the range of \pt\ probed, the jet yield on the away side remains roughly the same, but the per-trigger yield increases due to the decrease of the number of triggers at higher \pt; that property of $Y_{\pi\pi}^{\rm away}(p_t,p_b)$ is shown in Fig.\ 8. Since $\langle q'\rangle$ is insensitive to \pt, the shape of the \pb\ \dis\ should therefore also be insensitive to \pt, and that is confirmed by the property of $T_b$ in Fig.\ 9.

\begin{figure}[tbph]
\includegraphics[width=0.45\textwidth]{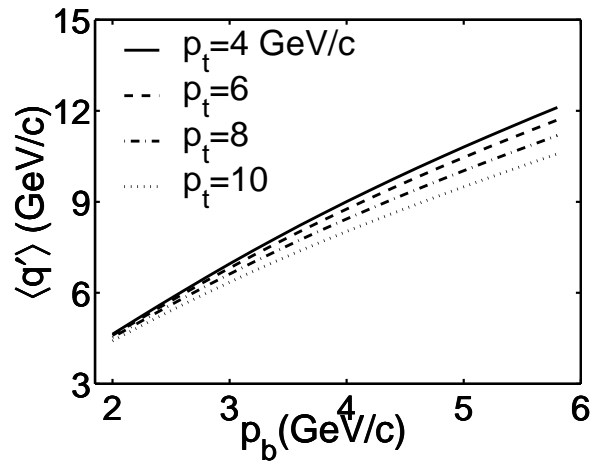}
\caption{Average value of parton momentum at the away-side surface for trigger momentum \pt\ and associated particle momentum \pb.}
\end{figure}

The values of $\langle q'\rangle$ are smaller than $\langle k'\rangle$ because of the longer path length for the recoil parton to reach the away side. We have already a hint of that in Fig.\ 10, since $\Gamma_{\rm away}(p_t, p_b)$ is also $\langle q' / k'\rangle(p_t,p_b)$, owing to the $\delta$ function in $G(q',k',L-t)$, despite the fact that $\langle q' / k'\rangle\ne \langle q'\rangle/\langle k'\rangle$ exactly. Nevertheless, at $p_t=p_b=4$ GeV/c, both $\langle q'\rangle/\langle k'\rangle$ and $\langle q' / k'\rangle$ are approximately 0.24.

Let us now make a more careful comparison between $\langle k\rangle$ and $\langle k'\rangle$, bearing in mind the difference in the  probability \dis s used in the averaging of $k$ and $k'$. If we set $\ptt$ in Fig.\ 6 to \pt\ in Fig.\ 11 and consider $p_b=p_t$ in the latter, then we find $\langle k'\rangle= 4.7 \langle k\rangle$ for $\ptt=p_t=p_b=4$ GeV/c, (dropping to 4.2$\langle k\rangle$ at $p_t=6$ GeV/c). Thus the condition of having an associated particle on the away side with equal momentum as the trigger eliminates the trigger bias and raises $\langle k'\rangle$ to approximately 9 \pt. The implication is that the average location of the hard-scattering point is in the middle of the medium (due to the symmetry of the two sides) and a large fraction of the parton energy is lost before exiting on either side. That fraction is 76\% as we have learned from $\langle q'\rangle/\langle k'\rangle$ above.

There are properties of the suppression factor that are noteworthy at the symmetry point $p_t=p_b$. We see in Fig.\ 10 that $\Gamma_{\rm away}(p_t,p_b)$ appears to be constant when $p_t=p_b$ is changed from 4 to 6 GeV/c. Let us then define
\begin{eqnarray}
\Gamma(p)=\Gamma_{\rm away}(p_t,p_b) , \qquad p=p_t=p_b  ,  \label{31}
\end{eqnarray}
and calculate its $p$ dependence. The result is shown in Fig.\ 13. It is indeed constant with the value $\Gamma(p)=0.24$. As we  have stated above, it follows from the $\delta$ function in $G(q',k',L-t)$ that  $\Gamma_{\rm away}(p_t,p_b)=\langle q'/k'\rangle(p_t,p_b)$. Thus for $p_t=p_b$, we have 
\begin{eqnarray}
\Gamma(p)=\langle q'/k'\rangle=\langle q/k\rangle=e^{-\beta L/2}  ,  \label{32}
\end{eqnarray}
the last equality being the consequence of identifying Eqs.\ (\ref{29}) and (\ref{30}) at the symmetry point. Putting $\beta L=2.9$ in Eq.\ (\ref{32}), one finds that $\Gamma(p)=0.235$, which is essentially the value 0.24 determined in Fig.\ 13. Thus we have consistency.

\begin{figure}[tbph]
\includegraphics[width=0.4\textwidth]{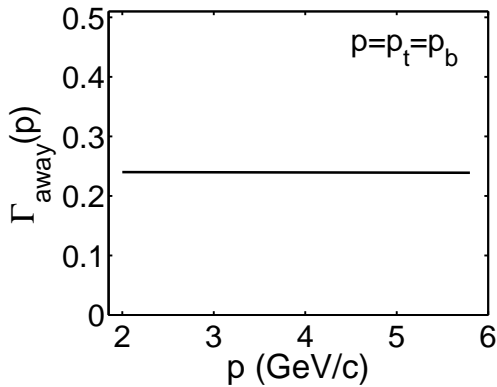}
\caption{Away-side suppression factor for symmetric momenta on the two sides: $p=p_t=p_b$.}
\end{figure}
\begin{figure}[tbph]
\includegraphics[width=0.45\textwidth]{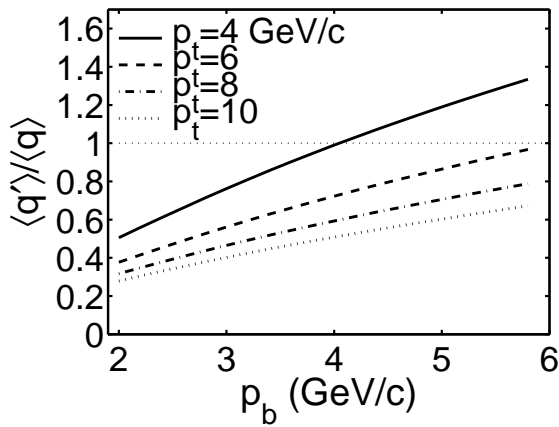}
\caption{The  ratio $\langle q'\rangle/\langle q\rangle$ as a function of \pt\ and \pb. It has the value 1 at $p_t=p_b$, shown explicitly at 4 and 6 GeV/c.}
\end{figure}

It is also of interest to compare $\langle q\rangle$ and $\langle q'\rangle$ for correlated particles on both sides. The result on $\langle q'\rangle/\langle q\rangle$ is shown in Fig.\ 14. Note that the ratio is 1 at the symmetry points $p_t=p_b$ at 4 and 6 GeV/c. For fixed \pt, the ratio increases with \pb, and, of course, for fixed \pb\ it decreases with \pt, since the inverse ratio, $\langle q\rangle/\langle q'\rangle$, should increase. This feature essentially describes how the hard-scattering point moves from one side of the midpoint to the other side, as \pb\ is changed from below \pt\ to above \pt. That is clearly a consequence of the counteracting effects of trigger and antitrigger biases.

\section{Centrality Dependence}

In the preceding section we have studied the properties of dihadron correlation on opposite sides for a nuclear slab of fixed $\beta L=2.9$, which is the average value of the quenching parameter in central collision. Before considering other centralities we must first establish a scheme to treat the realistic nuclear medium that has varying transverse width even for a fixed centrality, depending on the section of the overlap that the hard parton traverses. As a start we can formally treat $\bl$ as a variable and study how the yield depends on it.

\begin{figure}[tbph]
\includegraphics[width=0.4\textwidth]{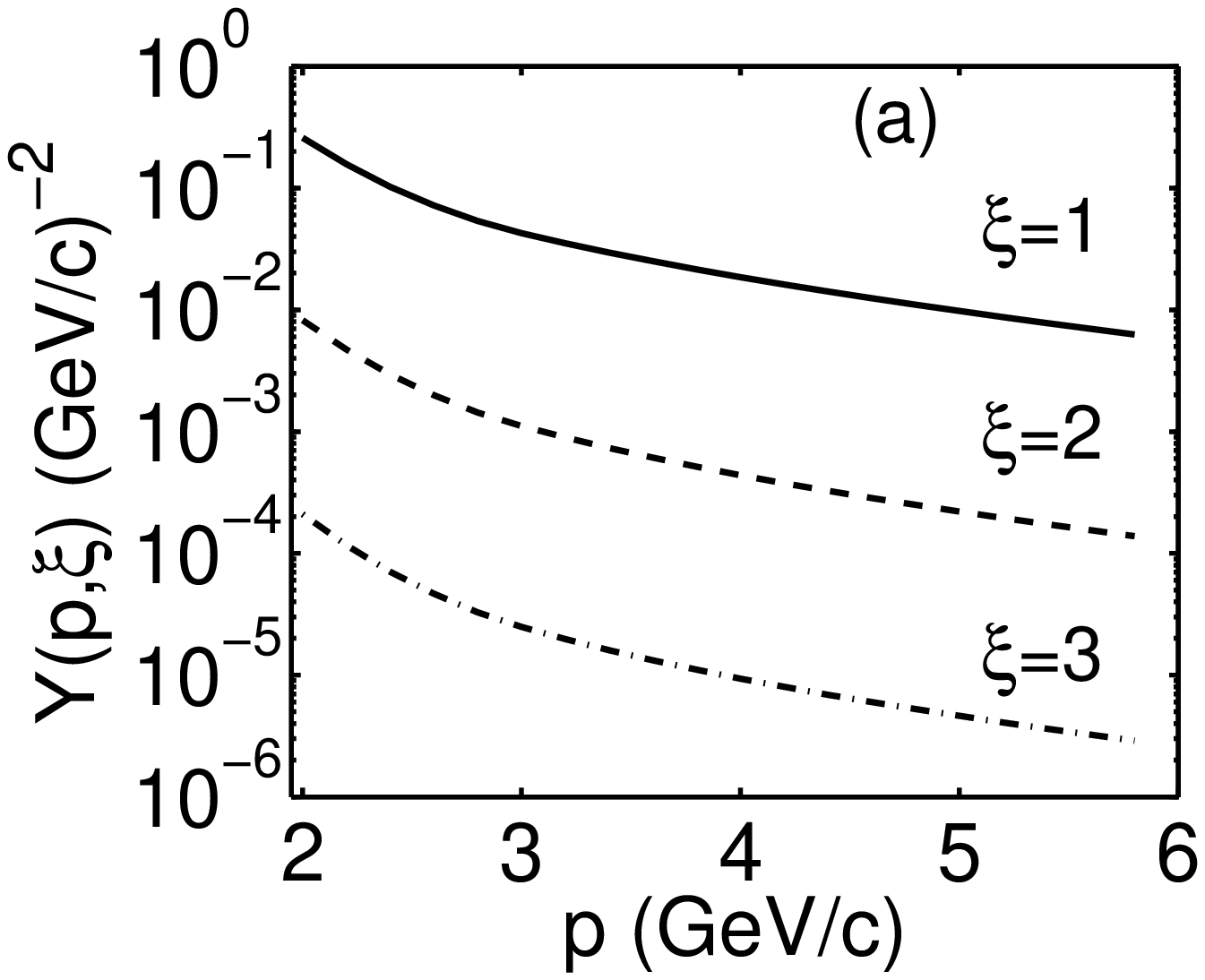}
\includegraphics[width=0.4\textwidth]{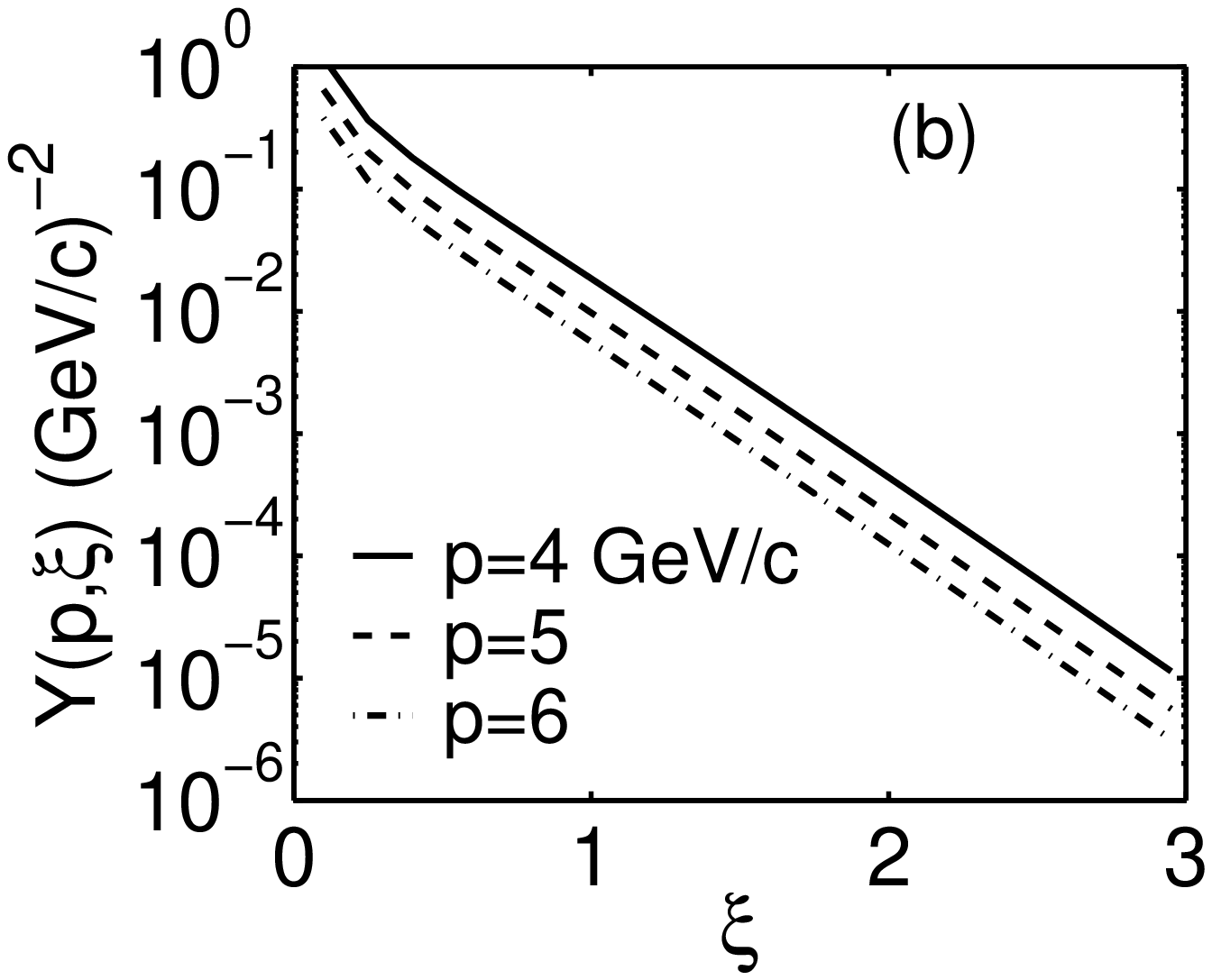}
\caption{Yield per trigger at the symmetry point $p=p_t=p_b$ for (a) fixed $\xi$, and (b) fixed $p$.}
\end{figure}

Let us define the per-trigger yield at the symmetry point
\begin{eqnarray}
Y(p,\xi)=Y_{\pi\pi}^{\rm away}(p_t,p_b,\bl) ,  \label{33}
\end{eqnarray}
where $p=p_t=p_b$ and $\xi=\bl$ now treated as a variable. Using Eqs.\ (\ref{24}) and (\ref{25}), we can calculate  $Y(p,\xi)$ with the results shown in Fig.\ 15(a) and (b) for fixed $\xi$ and $p$, respectively. The decrease with $p$ for fixed $\xi$ is more gentle than in Fig.\ 8 for unsymmetrical \pt\ and $p_b$. The decrease with $\xi$ for fixed $p$ is exponential for $\xi>0.5$, approximately as $e^{-3.8\xi}$, which is roughly what one expects from the $\bl$ dependence that one sees in \e (\ref{24}), remembering that $f_i(k)$ behaves in a power law as indicated in \e (\ref{26}) with $a\sim 7.7\div 8.7$. What we gain from Fig.\ 15 is a quantification of the picture that we already have in the increase of yield when the thickness of the nuclear medium is decreased. A corollary to that picture is that the hard parton momentum $k'$ need not be much larger than $p$, when $\xi$ is smaller. That is shown in Fig.\ 16, where $\langle k'\rangle$ decreases by nearly a factor of 3 when $\xi$ is decreased from 3 to 0.5.
\begin{figure}[tbph]
\includegraphics[width=0.4\textwidth]{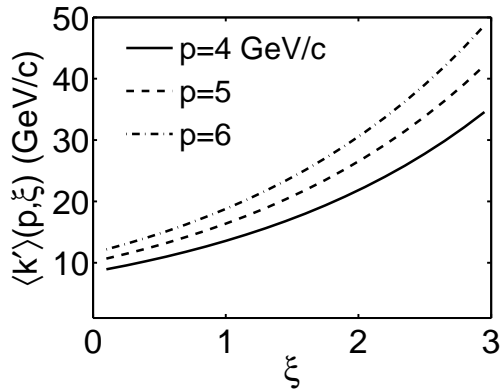}
\caption{Average value of recoil parton momentum as a function of $p$ and $\xi$.}
\end{figure}

The above consideration is merely a preview of what one should anticipate when we treat the medium realistically and  change the centrality. What we need first is a \dis\ of $\xi$ for a fixed centrality. Since quenching characteristic is involved in $\bl$, it is not just a geometrical problem of determining the path length in the elliptic overlap. The quenching effect depends on the local nuclear density and the location and orientation of the parton trajectory, so $\xi$ is a measure of the dynamical path length. Obviously, it is a very complicated problem for which no reliable solution is known. We shall approach it by first determining the average $\bl$ for single-particle inclusive \dis\ at each centrality $c$, as we have done in Fig.\ 2 for 0-10\% centrality. We use $c$ to denote the \% centrality so that $c=0.1$ means 10\% centrality, for example. We then construct a probability \dis\ $P(\xi,c)$ such that the average $\bar\xi(c)$ can fit the average $\bl$ as a function of $c$. With $P(\xi,c)$ at hand, it is then possible to calculate the yield per trigger for any centrality.

We start by revisiting Sec.\ 4 and the beginning of Sec.\ 5, but now consider all centralities $c$. For thermal partons the values of $C$ and $T$ in \e (4) as functions of $c$ are given in Ref.\ \cite{ht}. For hard partons, $f_i(k)$ is scaled by $N_{\rm coll}$. We use \e (\ref{18}) to calculate the single-pion $p_T$ \dis\ and obtain a good fit of the data, shown in Fig.\ 17, by adjusting $\bl(c)$ at each $c$. The data are from PHENIX \cite{phe} for $c=0.05, 0.15, \cdots, 0.86$ at intervals of 0.1. The fits are remarkably good for all centralities. The resulting $\bl(c)$ are shown by the solid dots in Fig.\ 18, which will serve as the key link between the centrality dependence of realistic nuclear collisions and the modeling of the quenching probability $P(\xi, c)$ at each $c$.

\begin{figure}[tbph]
\includegraphics[width=0.5\textwidth]{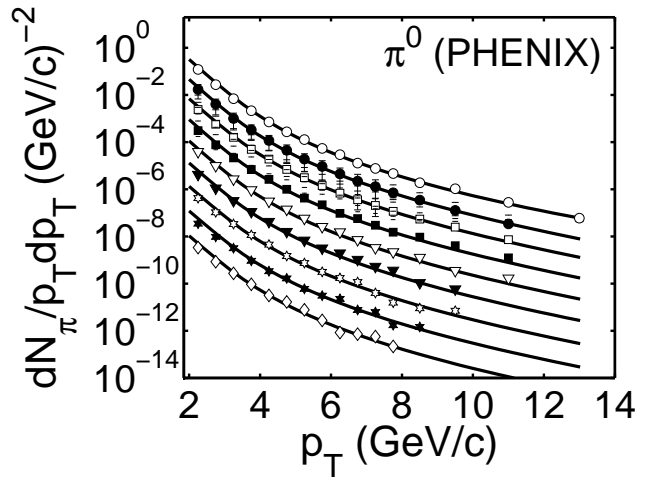}
\caption{Inclusive \dis\ of $\pi^0$ for all centralities ranging from 0-10\% (top) to 80-92\% (bottom) in 10\% steps, each displaced by a factor of 0.2. The data are from Ref.\ \cite{phe}. The curves are calculated from using \e (\ref{18}) with $\beta L(c)$ adjusted to fit and shown as dots in Fig.\ 18.}
\end{figure}
\begin{figure}[tbph]
\includegraphics[width=0.4\textwidth]{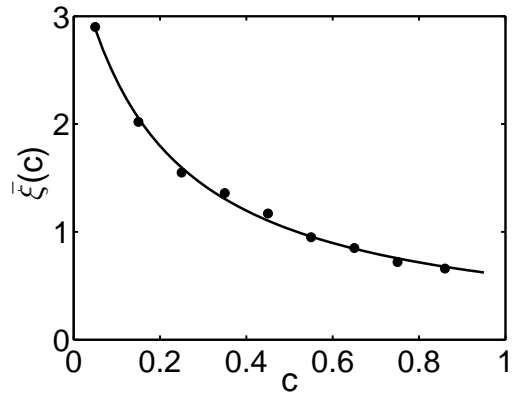}
\caption{The dots are the values of $\beta L(c)$ used to fit the inclusive \dis s in Fig.\ 17. The solid line is the average $\bar\xi(c)$ from the $\xi$ \dis\ $P(\xi, c)$ given in \e (\ref{34}).}
\end{figure}
\begin{figure}[tbph]
\includegraphics[width=0.4\textwidth]{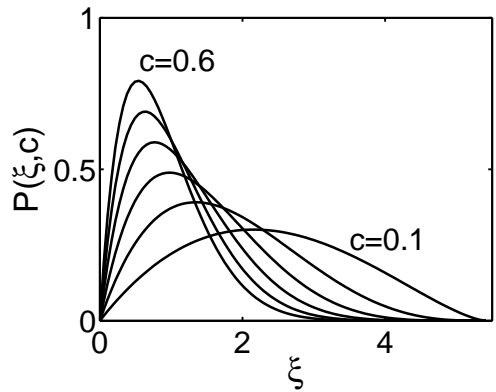}
\caption{The \dis\ of the dynamical path length $\xi$ for six values of centrality $c$ in steps of 0.1.}
\end{figure}

With the aim of fitting $\bl(c)$ in mind, it is sufficient to use a simple form for $P(\xi, c)$ that contains the basic features of noncentral collisions, namely: at any fixed $c$, $P(\xi, c)$ should have a maximum between the two ends of $\xi$, with the location of the maximum decreasing with increasing $c$. We adopt the form
\begin{eqnarray}
P(\xi, c)=N\xi(\xi_0-\xi)^{\alpha c},   \label{34}
\end{eqnarray}
where $N$ normalizes the total probabilty to 1, and $\xi_0$ and $\alpha$ are two parameters. We find that with 
\begin{eqnarray}
\xi_0=5.42,   \qquad  \alpha=15.2   \label{35}
\end{eqnarray}
we get the average $\bar\xi(c)$ that fits $\bl(c)$ very well, as shown by the solid line in Fig.\ 18. The \dis\ $P(\xi, c)$ itself is shown in Fig.\ 19 that exhibits the decrease of the maxium with increasing $c$. In view of the difficulty of deriving $\bl(c)$ from first principles, let alone $P(\xi,c)$, we regard  Eqs.\ (\ref{34}) and (\ref{35}) as being totally satisfactory for the description of how the path-dependent quenching parameter varies among the collisions within each class of centrality.

With $P(\xi,c)$ thus obtained, we can now return to the per-trigger yield $Y(p,\xi)$ at the symmetry point defined in Eq.\ (\ref{33}). To determine the yield at a fixed centrality it is not simply a matter of averaging $Y(p,\xi)$ over all $\xi$, using $P(\xi,c)$ as the weighting factor at each $\xi$.  $Y(p,\xi)$ that is shown in Fig.\ 15 is obtained for centrality being held at $c=0.05$, while $\xi$ is varied. We must redo the calculation for $Y_{\pi\pi}^{\rm away}(p_t,p_b)$ using \e (\ref{24}) and (\ref{25}), but now include also the dependencies of $C, T$ and $f_i(k)$ on $c$, discussed above. That is, we define 
\begin{eqnarray}
{dN_{\pi}(c)\over p_tdp_t}&=&\int d\xi P(\xi,c){dN_{\pi}(c,\xi)\over p_tdp_t},  \label{36} \\
{dN_{\pi\pi}(c)\over p_tp_bdp_tdp_b}&=&\int d\xi P(\xi,c){dN_{\pi\pi}(c,\xi)\over p_tp_bdp_tdp_b},  \label{37} \\
Y_{\pi\pi}^{\rm away}(p_t,p_b,c)&=&{dN_{\pi\pi}(c)\over p_tp_bdp_tdp_b}\left/ {dN_{\pi}(c)\over p_tdp_t}\right. ,  \label{38}
\end{eqnarray}
and calculate
\begin{eqnarray}
Y(p,c)=Y_{\pi\pi}^{\rm away}(p=p_t=p_b,c) .   \label{39}
\end{eqnarray}
The results are shown in Fig.\ 20 for four centralities. Note that
the dependence on $c$ is not as drastic as the dependence of $Y(p,\xi)$ on $\xi$ in Fig.\ 15(a), which shows the dominance of $\xi=1$ over $\xi=3$, so upon averaging over $\xi$ at each $c$ the small $\xi$ contribution is always more important at any $c$. The per-trigger yield rises with $c$ because of reduced medium suppression. The $p$ dependence appears to be universal.

\begin{figure}[tbph]
\includegraphics[width=0.45\textwidth]{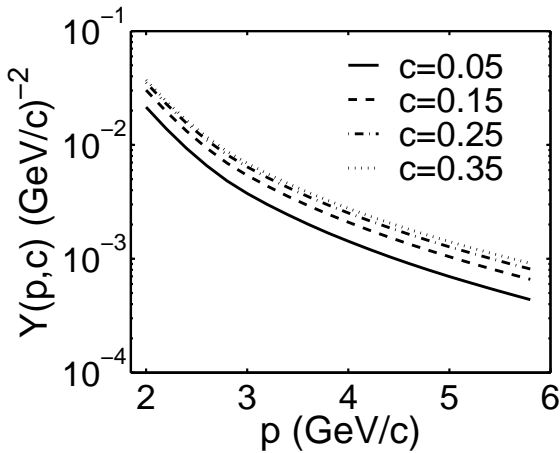}
\caption{Yield per trigger at the symmetric point $p=p_t=p_b$ for four values of centrality.}
\end{figure}
\begin{figure}[tbph]
\includegraphics[width=0.45\textwidth]{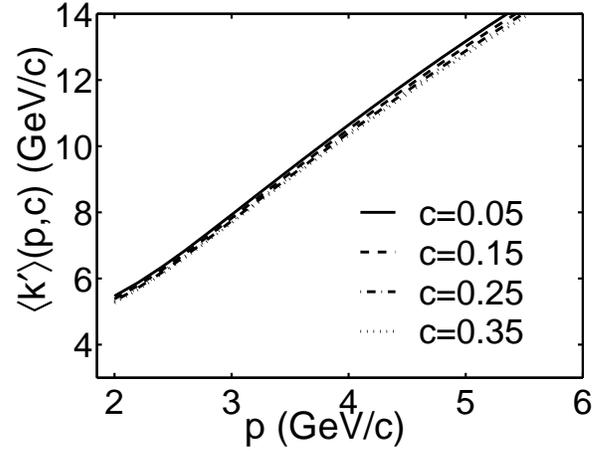}
\caption{Average value of the initial hard parton momentum directed at the away side for hadron momenta $p=p_t=p_b$ for four values of centrality.}
\end{figure}
\begin{figure}[tbph]
\includegraphics[width=0.45\textwidth]{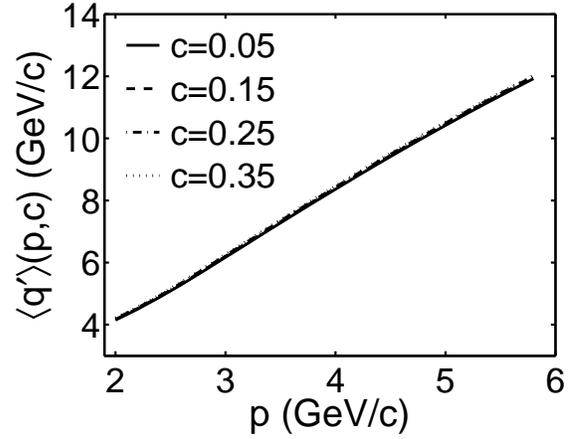}
\caption{Average value of the  parton momentum  at the away-side surface for hadron momenta $p=p_t=p_b$ for four values of centrality.}
\end{figure}

Related to the mild dependence on $c$ in Fig.\ 20, we can investigate the properties of $\langle k'\rangle$. Recall from Fig.\ 11 that for fixed $\bl=2.9$, $\langle k'\rangle$ is much larger than \pt\ or \pb, the phenomenon referred to as a feature of  antitrigger bias. We have also seen that at the symmetry point $p=p_t=p_b$ the values of $\langle k'\rangle(p,\xi)$ decrease significantly at lower $\xi$, shown in Fig.\ 16.
To calculate $\langle k'\rangle(p,c)$ for different $c$ we again cannot simply average $\langle k'\rangle(p,\xi)$ over $\xi$ using $P(\xi,c)$ as weight, since the normalization factor of $\langle k'\rangle(p,\xi)$ must also be averaged over $\xi$ separately. The result for $\langle k'\rangle(p,c)$, shown in Fig.\ 21,  exhibits essentially no dependence on $c$. The magnitude is approximately $2.5\,p$, which is much lower than $\langle k'\rangle(p,\xi)$ in Fig.\ 16 and more like the near-side $\langle k\rangle(\ptt)$ in Fig.\ 6. Furthermore, $\langle q'\rangle(p,c)$ can also be calculated in the same manner with  similar result shown in Fig.\ 22. The small difference is that whereas $\langle k'\rangle(p,c)$ decreases slightly with $c$, $\langle q'\rangle(p,c)$ increases imperceptibly. The ratio $\langle q'\rangle/\langle k'\rangle$ is seen in Fig.\ 23 to be nearly constant at around 0.8, increasing about 4\% when $c$ changes from 0.05 to 0.35. The value of that ratio is roughly the same as the value of $\Gamma_{\rm near}(\ptt)$ at comparable $\ptt$ in Fig.\ 5, which corresponds to $\langle q/k\rangle$ on the near side  without the requirement of a recoil jet. Thus the medium degrades  the parton momentum from $k'$ to $q'$ on the away side by about the same degree as from $k$ to $q$ on the near side, and the degree of suppression is essentially independent of centrality. The inescapable conclusion is then that when symmetric back-to-back hadron momenta $(p_t=p_b)$ are required, the dijets that give rise to them are due to hard partons created very near the surface on both sides so that they  suffer minimal energy loss as they propagate in opposite directions through the rim of the nuclear medium. That means they must be tangential jets. This is a remarkable result that emerges from the calculation, and is consistent with the dijet+1 correlation data \cite{ob} in which no ridge is found and whose $N_{\rm part}^{2/3}$ dependence suggests that they are generated near the surface, i.e., tangential jets.

\begin{figure}[tbph]
\includegraphics[width=0.45\textwidth]{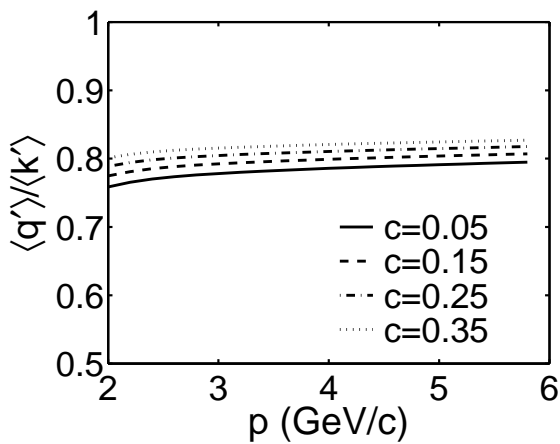}
\caption{The ratio $\langle q'\rangle/\langle k'\rangle$ at the symmetry point $p=p_t=p_b$ for four values of centrality.}
\end{figure}

\section{Yields at Unequal Trigger and Associated Particle Momenta}

Having studied in the previous section on how the yield at the symmetry point $p=p_t=p_b$ behaves at different centralities, we can finally investigate the properties at asymmetric points and appreciate the significance of small variations. We first consider the per-trigger yield of an associated particle on the near side at centrality $c$. The basic formula is as in \e (\ref{21}), except that both the numerator and denominator must be averaged over $P(\xi,c)$ separately, similar to Eqs.\ (\ref{36})-(\ref{38}). 
The centrality dependence of the result is shown in Fig.\ 24 for $p_t=4$ and 6 GeV/c and $p_a=2$ and 4 GeV/c. The near-side yield is nearly constant in $c$, and decreases with \pa\ for a fixed \pt, but increases with \pt\ for a fixed \pa. The solid lines in that figure represent the integrated results for $2<p_a<4$ GeV/c. 
 The data in  Fig.\ 24 are  for $3<p_T^{\rm trig} <4$ GeV/c and $p_T^{\rm assoc}>2$ GeV/c  \cite{put}. The agreement between our result and the data is remarkably good both in magnitude and in $c$ dependence. The magnitude of the integrated yield is sensitively dependent on the lower limit of integration in \pa, so other data on centrality dependence with different lower limits, such as in \cite{jb2,cn}, cannot be compared with the black line in Fig.\ 24, although the rough insensitivity to $N_{\rm part}$ is seen irrespective of the cut in $p_T^{\rm assoc}$.
 The approximate independence on centrality is a manifestation of the trigger bias, as we have already noted in Sec. 4 that the hard-scattering point is in a layer roughly 13\% of $L$ inside the near-side surface and is insensitive to how large the main body of the medium is. However, the thermal and shower partons have different dependencies on $c$ and the decrease of TS recombination with increasing $c$ cancels the increase of SS recombination with $c$ so that their sum results in approximate independence on $c$.

\begin{figure}[tbph]
\includegraphics[width=0.5\textwidth]{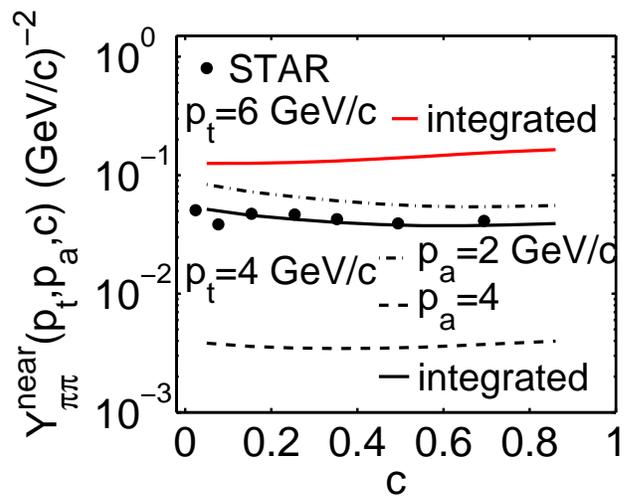}
\caption{(Color online) Yield per trigger in the near-side jet as functions of centrality $c$ for $p_t=4$ GeV/c in black lines and $p_t=6$ GeV/c in red line. Dash-dotted line is for $p_a=2$ GeV/c and dashed line for $p_a=4$ GeV/c. The solid lines are for the yields integrated over \pa\ from 2 to 4 GeV/c. The data are from Ref.\ \cite{put} for $3<p_T^{\rm trig}<4$ GeV/c and $p_T^{\rm assoc}>2$ GeV/c.}
\end{figure}
\begin{figure}[tbph]
\includegraphics[width=0.45\textwidth]{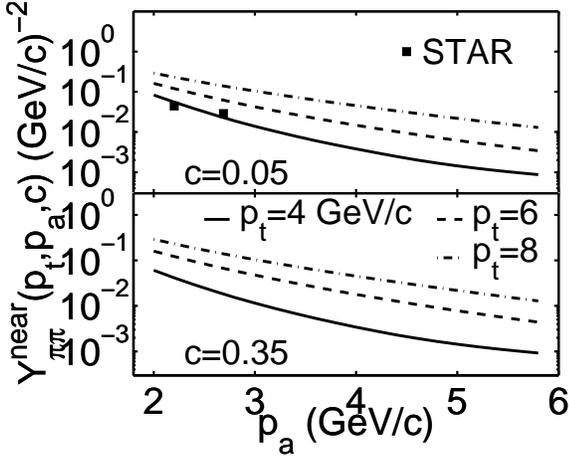}
\caption{Yield per trigger in the near-side jet as functions of \pa\  for two values of $c$ and three values of \pt. Data points are from Ref.\ \cite{cn}; see text for details.}
\end{figure}

To see the dependence on \pa\ for fixed \pt\ we show the yield in Fig.\ 25 for two representative values of $c$ at 0.05 and 0.35.  These \dis s are very similar to $Y_{\pi\pi}^{\rm near}$  in Fig.\ 3, which is the yield for fixed $\beta L=2.9$. Thus the result is the same whether we fix $c$ or $\bl$. The inverse slope $T_a$ is therefore essentially what is shown in Fig.\ 4 already. In Fig.\ 25 we have included two data points from Ref.\ \cite{cn}, where recent results on near-side correlations have been reported. The data for central Au+Au collisions (0-10\%) at 200 GeV are given for integrated jet yield per trigger (for $-1<\Delta\phi<1$) with $3<p_T^{\rm trig}<6$ GeV/c and $1.5<p_T^{\rm assoc}<p_T^{\rm trig}$. Since our calculation is for per-trigger yield of particles in the near-side jet averaged over all $\Delta\phi$ in the jet, we have divided the data by 2 (the range of $\Delta\phi$), and include only the  points at $p_T^{\rm assoc}>2$ GeV/c in the upper panel of Fig.\ 25. Our curve for $p_t=4$ GeV/c agrees very well with those two data points, which are averaged over the range $3<p_t<6$ GeV/c.

\begin{figure}[tbph]
\hspace*{-.2cm}
\includegraphics[width=0.51\textwidth]{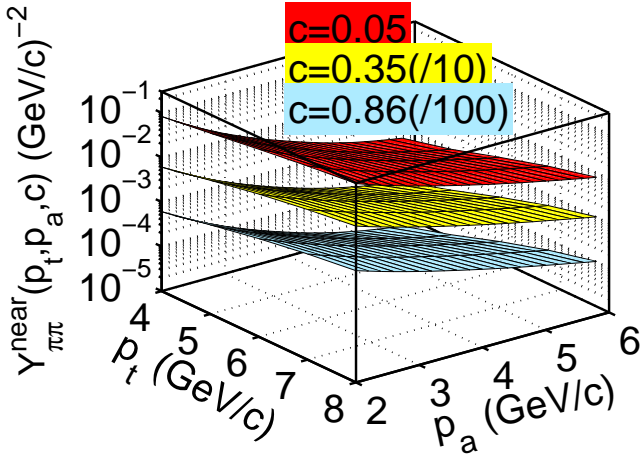}
\caption{(Color online) A 3D plot of $Y_{\pi\pi}^{\rm near}(p_t,p_a,c)$ for $c=0.05$ (red), $c=0.35$ (yellow) lowered by a factor of $10^{-1}$, and  $c=0.86$ (blue) lowered by  $10^{-2}$.}
\end{figure}

An overall view of $Y_{\pi\pi}^{\rm near}(p_t,p_a,c)$ as a function of both \pt\ and \pa\ for three illustrative values of $c$ is shown in Fig.\ 26. For clarity's sake we have multiplied the yield for $c=0.35$ (in yellow) by $10^{-1}$ and for $c=0.86$ (in blue) by $10^{-2}$. The increase with \pt\ is perceptible, while the dependence on $c$ is negligible.

For the away-side yield we use \e (\ref{38}) and obtain the results shown in Fig.\ 27, where a factor of about 2 increase in the magnitude is seen when $c$ is raised from 0.05 to 0.35. Thus when the nuclear overlap is smaller, it is easier for the recoil jet to reach the away side and to produce a particle at $p_b$. The shape of the \pb\ \dis\ is basically independent of centrality, since the hadronization process does not change with $c$. Figure 28 shows a 3D plot of $Y_{\pi\pi}^{\rm away}(p_t,p_b,c)$, again with $c=0.35$ (in yellow) and  $c=0.86$ (in blue) lowered by factors of $10^{-1}$ and $10^{-2}$, respectively. The near independence on \pt\ is evident, while the increase with $c$ is only from 0.05 to 0.35, but not from 0.35 to 0.86.

\begin{figure}[tbph]
\includegraphics[width=0.45\textwidth]{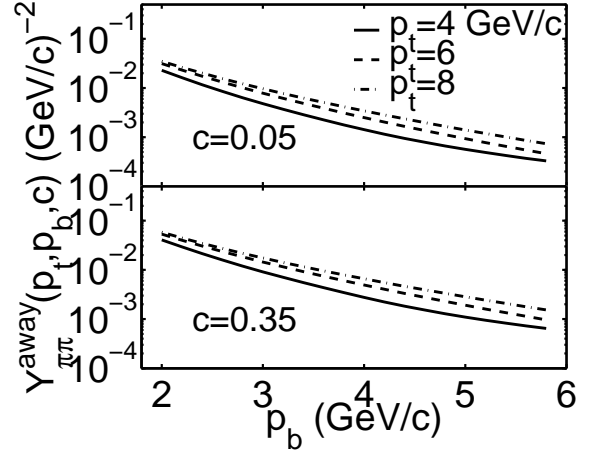}
\caption{Yield per trigger in the away-side jet plotted in the same format as in Fig.\ 25 with \pa\ replaced by \pb.}
\end{figure}
\begin{figure}[tbph]
\hspace*{-.2cm}
\includegraphics[width=0.51\textwidth]{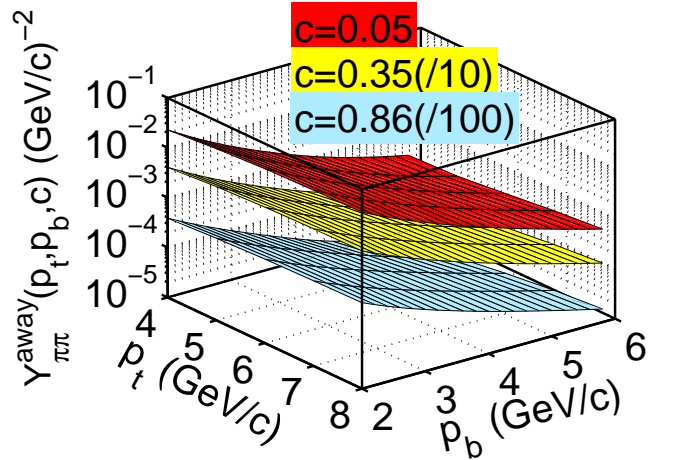}
\caption{(Color online) Same as in Fig.\ 26 but for $Y_{\pi\pi}^{\rm away}(p_t,p_b,c)$.}
\end{figure}

For fixed $c$ and varying combinations of \pt\ and \pb, we can determine a quantitative measure of the antitrigger bias by calculating the average $\langle \beta t'\rangle$, where $t'$ denotes the distance from the hard-scattering point to the away-side surface. In the calculation we identify $\beta t'$ with $\ln (k/q')$ by virtue of $G(q',k,L-t)$ in \e (\ref{22}).
Figure 29 shows the results for $c=0.05$ and 0.35. For fixed \pt, $\langle \beta t'\rangle$ decreases with increasing \pb\ as the scattering point is pulled closer to the away side. For fixed \pb, that point moves closer to the near side, as \pt\ increases, thus increasing $\langle \beta t'\rangle$. The whole set of curves are lower at higher $c$. Thus Fig.\ 29 provides a good description of the antitrigger bias. Note that the magnitude of $\langle \beta t'\rangle$ is not large, less than 0.5 even for $c=0.05$. It is much smaller than the value $\beta L=2.9$ in \e (\ref{28}), which is for the single-particle inclusive \dis. Again, we see that when a particle on the away side is required, the scattering point cannot be too far from the surface of the away side. At $p_t=p_b=4$ GeV/c, we have $\langle \beta t'\rangle\approx 0.2$ for both values of $c$. That is just the value of $\langle \beta t\rangle$ at $\ptt=4$ GeV/c in Fig.\ 7, consistent with the scattering point being midway between the two sides. As we have learned from the preceding section, when $p_t=p_b$ the two jets produced are tangential jets near the rim of the overlap. As \pt\ is increased, the scattering point can be embedded deeper in the interior, so $\langle \beta t'\rangle$ increases, but not very much.

\begin{figure}[tbph]
\includegraphics[width=0.45\textwidth]{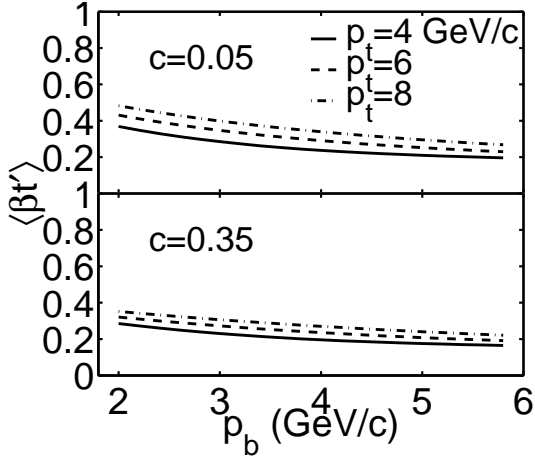}
\caption{Average value of $\beta t'$ where $t'$ is the distance between the hard-scattering point and the away-side surface for various values of $c, p_t$ and \pb.}
\end{figure}

We can also present a more explicit description of the antitrigger bias in terms of parton momenta. We show in Fig.\ 30 the average momentum $\langle k'\rangle(p_t,p_b,c)$ of the hard parton directed away from the trigger as a function of \pb\ for fixed $c$ and \pt. The increase with \pb\ is now much slower than the increases with $p\  (=p_t=p_b)$ in Fig.\ 21. The difference between $c=0.05$ and 0.35 is minor, as in Fig.\ 21. Focusing on $p_t=4$ GeV/c, we see that $\langle k'\rangle$ increases from $\sim 8$ to $\sim 13$ GeV/c, as \pb\ increases  from 2 to 6 GeV/c, the magnitude being significantly lower than the corresponding $\langle k'\rangle$ in Fig.\ 11 for fixed $ \beta L=2.9$. However, compared to $\langle k\rangle$ on the near side in Fig.\ 6, where $\langle k\rangle\approx 8$ GeV/c at $p_T=4$ GeV/c, $\langle k'\rangle$ starts from about the same value at low \pb\ but increasing persistently with \pb, although \pt\ is fixed. That is, $\langle k'\rangle$ is always greater than $\langle k\rangle$
at fixed $c$ and \pt\ for one of the following two reasons.  If \pb\ is less than \pt, then   the longer path length on the away side due to antitrigger bias leads to higher $\langle k'\rangle$ despite the momentum balance $k'=k$ in every hard scattering event. If \pb\ is more than \pt, then clearly the jet momentum on the away side must on average be higher than on the trigger side.

\begin{figure}[tbph]
\includegraphics[width=0.45\textwidth]{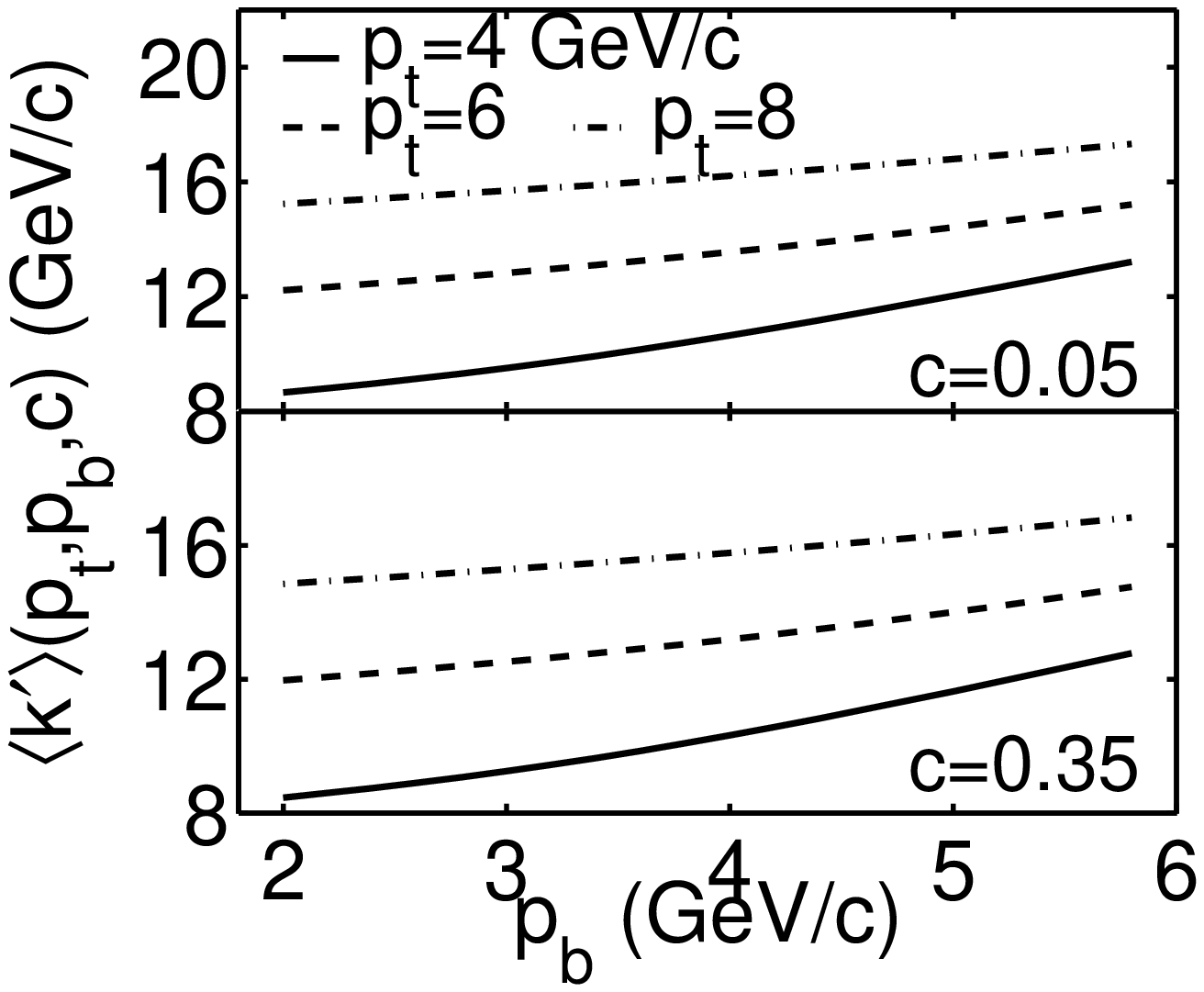}
\caption{Average value of the initial hard parton momentum directed at the away side for various hadron momenta \pt\ and \pb\ for two values of centrality.}
\end{figure}
\begin{figure}[tbph]
\includegraphics[width=0.45\textwidth]{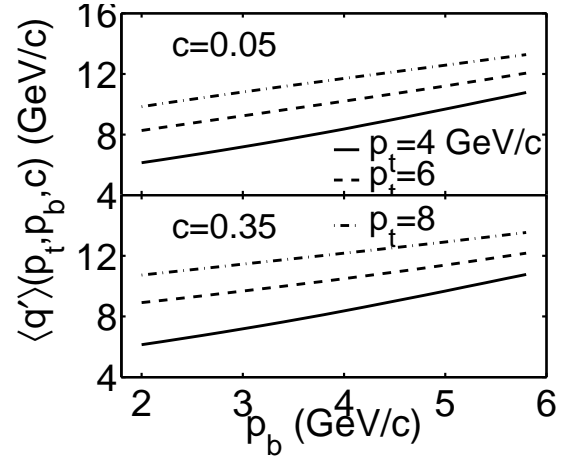}
\caption{Average value of the   parton momentum  at the away side surface for various \pt, \pb, and $c$ as in Fig.\ 30.}
\end{figure}

The behavior of $\langle q'\rangle(p_t,p_b,c)$ at the away-side surface, shown in Fig.\ 31, differs from that in Fig.\ 22 in the same way that Fig.\ 30 differs from Fig.\ 21. Compared to $\langle k'\rangle(p_t,p_b,c)$, the magnitude of $\langle q'\rangle(p_t,p_b,c)$ are, of course, lower, but the \pt\ and \pb\ dependencies are similar. More revealing is the ratio $\langle q'\rangle/\langle k'\rangle(p_t,p_b,c)$ in Fig.\ 32, which shows the effect of energy loss that decreases (difference from 1) with increasing \pb\ due to decreasing path length, but increases with increasing \pt\ due to increasing path length. The push-and-pull effect of \pt\ and \pb\ is now clearly depicted in Figs.\ 30-32 that could not be shown in Figs.\ 21-23 where $p=p_t=p_b$.

\begin{figure}[tbph]
\includegraphics[width=0.45\textwidth]{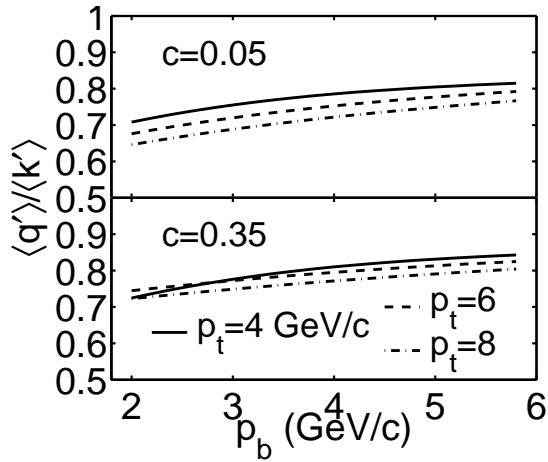}
\caption{The ratio  $\langle q'\rangle/\langle k'\rangle$ for \pt, \pb\ and $c$ as in Figs.\ 30 and 31.}
\end{figure}

We have been able to exhibit these characteristics of the medium effects by calculating theoretical quantities, such as $\langle \beta t'\rangle, \langle k'\rangle$ and $\langle q'\rangle$, for various values of \pt, \pb\ and $c$ that are experimentally measurable. Beside the per-trigger yields $Y_{\pi\pi}^{\rm near}(p_t,p_a,c)$ and $Y_{\pi\pi}^{\rm away}(p_t,p_b,c)$ that can be directly checked by experiment, it is possible that $\langle q\rangle$ and $\langle q'\rangle$ can indirectly be estimated by studying the total transverse momenta of charged particles in the near- and away-side jets.

\section{Conclusion}

We have made an exhaustive investigation of the properties of dihadron correlation in jets produced in heavy-ion collisions. Our treatment is based on a reliable description of hadronization through recombination on one hand, and on a realistic accounting of the medium effect on the other. Only two free parameters are used to specify the \dis\ of dynamical path length $\xi$ for any centrality, and they are determined by fitting over 100 data points of the $\pi^0$ inclusive spectra. All other parameters have been fixed by previous work in the recombination model. Thus our study of the hadronic correlation with high-$p_T$ trigger has very little freedom for adjustment, and for that reason we have been able to calculate unambiguously many quantities that reveal the medium effect on partons and the relationship among the momenta of hadrons that they produce. Not all the quantities calculated can be measured, but among those that can be checked by experiments, each encounter with existing data exhibits good agreement. It is therefore reasonable to conclude that the theoretical framework presented here offers a reliable description of one- and two-jet production at RHIC energy.

One outcome of this study is the determination of the probability $P(\xi,c)$ of a hard parton having a dynamical path length $\xi$ in a collision at centrality $c$, where $\xi$ plays the role of the suppression parameter $\beta L$, except that it varies among all possible trajectories and density-dependent energy-loss factors, while $\bl$ is the average over all $\xi$. The behavior of $P(\xi,c)$ exhibited in Fig.\ 19 for various $c$ may be regarded as the fruit of the program to learn about the medium effect from heavy-ion collisions. Our calculation of the dihadron correlation in jets for each centrality would not have been reliable enough to compare favorably with data as in Figs.\ 24 and 25, if we did not have $P(\xi,c)$ to link theory with experiment.

Our study has shed light on the trigger and antitrigger biases, which are brief terms, referring to the complex issues involving nuclear geometry and medium suppression, that we have made more precise by examining the average momenta $\langle k\rangle,\langle k'\rangle,\langle q\rangle$ and $\langle q'\rangle$ under different conditions. The ratios of those quantities reveal the different suppression factors on partons propagating toward the near and away sides. When the trigger momentum and the away-side associated-particle momentum are equal, we have learned, somewhat by surprise, that the yield is dominated by tangential jets, essentially independent of centrality. Revealing results such as that await precise verification by experiments. 

We have restricted our attention in this paper to the dependence on $\ptt$ only, and to the range of $\ptt$ large enough to leave out the consideration of ridge on the near side and of double humps on the away side. Ridgeology is a separate subject in its own right and is treated elsewhere \cite{ch3}.  Lowering the $\ptt$ range would contaminate the trigger with medium partons through thermal-thermal recombination, a situation well within the capability of our formalism to handle, but not considered here. Our treatment can also be generalized to include the azimuthal angle $\phi$. The dependence of jet production on the trigger angle $\phi_s$ relative to the reaction plane would be very interesting to study, as has already been initiated in data analysis by STAR \cite{af}. The widths of the jet peaks on the two sides, as well as the possible misalignment of the back-to-back jets, are challenging problems still to be investigated.

A significant qualitative conclusion that can be drawn from the many ways of posing our questions on the medium effect is that it is hard experimentally to probe the interior of the collision zone. The profile of the nuclear overlap is like a small island whose inhabitants may live uniformly  throughout the island, but only the ones near the shore can react easily to activities in the sea. The detected trigger jets are produced mainly along the near side, like the coastal inhabitants responding to a call from the sea. If an away-side jet is detected at the same time as the trigger jet, the production point is moved along the rim so that the back-to-back jets are dominantly tangential jets, just as in the analogy where the inhabitants that can respond to calls from both sides are the ones on the coast with unobstructed vision of the two sides.  In that sense jet tomography fails to probe the interior of medium, since rim production overwhelms any signal arising from the interior in one- and two-jet events. The situation naturally suggests that three-jet events may reach the interior, analogous to the inland dwellers of the island being able to have equal, though harder, access to all points at sea. In $e^+e^-$ annihilation gluon jets were discovered in events where three jets were produced, each being at nearly equal azimuthal angles from the other two. Similarly, three jets originating from a common point in heavy-ion collision are possible, such as in the gluonic process $g+g\to g+g+g$. They are, of course, less abundant than two jets for the dual  reason of higher order and enhanced suppression. However, the more serious experimental difficulty is to distinguish the 3-jet events from the background  consisting of the double-hump structure in the conventional away-side $\Delta\phi$ \dis. The latter is due to TT recombination, if the double humps can be related to the Mach cone, whereas the 3-jet event structure is associated with shower partons in each jet. The subject is rich and worthy of attention, if the medium interior is to be probed by jet tomography.

Another area of extension from this work is obviously heavy quark physics. In our sum over all parton types $i$ in Eqs.\ (\ref{18}), (\ref{19}) and (\ref{23}), we have limited to $g, u, d$ and $s$, and their antiquarks. To consider charm quark, for example, we must redo everything from the beginning, including the shower parton \dis\ and single-hadron ($D$) spectrum. The basic formalism is, however, the same as we have given here. A new $P(\xi,c)$ would have to be found for heavy quarks, and different possibilities of dijet correlation may reveal different medium effects \cite{rh}.

Finally, it is worth commenting that dijet correlation will undoubtedly be drastically different at LHC where not only numerous jets with $\ptt<20$ GeV/c will be created, but also the recombination of shower partons from neighboring jets can totally change the correlation between hadrons \cite{hy5}. The formalism used in this work will have to be thoroughly revised. Even the notion of jets distinguishable from background will require reexamination. The challenging work ahead will, however, not be daunting but stimulating, both theoretically and experimentally.

\section*{Acknowledgment}
We have been enlightened by communication with J.\ Bielcikova, C.\-H.\ Chen, E.\ Elhalhuli, B.\ Jacak, J.\ Jia, C.\ Nattrass and A.\ Sickles. This work was supported  in part,  by the U.\ S.\ Department of Energy under Grant No. DE-FG02-96ER40972 and by the National Natural Science Foundation of China under Grant No. 10635020 and 10775057 and by the Ministry of Education of China under project IRT0624.

\end{document}